# Probing the Meta-Stability of Oxide Core/Shell Nanoparticle Systems at Atomic Resolution

*Manuel A. Roldan,[a,b,1,‡] Arnaud Mayence,[c,‡] Alberto López-Ortega,[d] Ryo Ishikawa,[e] Juan Salafranca,[a] Marta Estrader,[f,g] German Salazar-Alvarez,[h,2] M. Dolors Baró,[i] Josep Nogués,[g,j,*] Stephen J. Pennycook,[k] Maria Varela[a,1,*]*

[a]Departamento de Física de Materiales & Instituto Pluridisciplinar, Universidad Complutense de Madrid, 28040 Madrid, Spain.
[b]John M. Cowley Center for High-Resolution Electron Microscopy, Arizona State University, 85231 AZ USA
[c]Department of Materials and Environmental Chemistry, Arrhenius Laboratory, Stockholm University, 10691 Stockholm, Sweden
[d]Instituto de Nanociencia, Nanotecnología y Materiales Moleculares and Depto. de Física Aplicada, Universidad de Castilla-La Mancha, 45071 Toledo, Spain
[e]Institute of Engineering Innovation, University of Tokyo, Tokyo 113-8656, Japan
[f]Departament de Química Inorgànica i Orgànica, Universitat de Barcelona, 08028 Barcelona, Spain
[g]Catalan Institute of Nanoscience and Nanotechnology (ICN2), CSIC and BIST, Campus UAB, Bellaterra, 08193 Barcelona, Spain.
[h]Department of Materials Science and Engineering, Ångström Laboratory, Uppsala University. SE-751 21 Uppsala, Sweden.
[i]Departament de Física, Universitat Autònoma de Barcelona, E-08193 Bellaterra (Barcelona), Spain.
[j]ICREA, Pg. Lluís Companys 23, E-08010 Barcelona, Spain.
[k]Department of Materials Science & Engineering, National University of Singapore, Singapore 117575.

[‡] *M. A. Roldan and A. Mayence contributed equally to the work.*
[1] *Previous Address: Materials Science & Technology Division, Oak Ridge National Laboratory, Oak Ridge TN 37831, USA.*
[2] *Previous address: Department of Materials and Environmental Chemistry, Arrhenius Laboratory, Stockholm University, 10691 Stockholm, Sweden.*

*Corresponding authors:*
*Prof. Dr. María Varela. mvarela@ucm.es*
*Prof. Dr. Josep Nogues. josep.nogues@icn2.cat*

**ABSTRACT:** Hybrid nanoparticles allow exploiting the interplay of confinement, proximity between different materials and interfacial effects. However, to harness their properties an in-depth understanding of their (meta)stability and interfacial characteristics is crucial. This is especially the case of nanosystems based on functional oxides working under reducing conditions, which may severely impact their properties.

In this work, the in-situ electron-induced selective reduction of $Mn_3O_4$ to MnO is studied in magnetic $Fe_3O_4/Mn_3O_4$ and $Mn_3O_4/Fe_3O_4$ core/shell nanoparticles by means of high-resolution scanning transmission electron microscopy combined with electron energy-loss spectroscopy. Such in-situ transformation allows mimicking the actual processes in *operando* environments. A multi-stage image analysis using geometric phase analysis combined with particle image velocity enables direct monitoring of the relationship between structure, chemical composition and strain relaxation during the $Mn_3O_4$ reduction. In the case of $Fe_3O_4/Mn_3O_4$ core/shell the transformation occurs smoothly without the formation of defects. However, for the inverse $Mn_3O_4/Fe_3O_4$ core/shell configuration the electron beam-induced transformation occurs in different stages that include redox reactions and void formation followed by strain field relaxation via formation of defects. This study highlights the relevance of understanding the local dynamics responsible for changes in the particle composition in order to control stability and, ultimately, macroscopic functionality.

- 

**Highlights**

- In situ, atomic resolution, electron microscopy study of the electron beam induced Mn3O4 → MnO reduction reaction in $Fe_3O_4$/$Mn_3O_4$ and $Mn_3O_4$/$Fe_3O_4$ core/shell nanoparticles.

- In depth analysis of the evolution of the microstructure with complex analysis techniques, including particle image velocimetry analysis and geometric phase analysis

- Very different evolution of the reaction depending on the morphology, i.e., $Mn_3O_4$ in the core or in the shell.

- Multistage complex processes taking place in the reduction process in the $Mn_3O_4$/$Fe_3O_4$ with generation of voids, stress relaxation, ionic transport and formation of defects.

# 1. Introduction

Hybrid nanoparticles [1] are very attractive as model systems for fundamental studies [2-6] and also for a very wide range of applications. The combination of reduced dimensionality effects with the interplay between the properties of the constituents can give rise to multiple, and sometimes new, functionalities in systems of interest for catalytical [7], optical [8,9] magnetic [10,11] or biomedical applications [12-15]. Such multifunctional nanoparticles can be obtained in different configurations, being the dimer and core/shell geometries the most usual archetypes [1]. Interestingly, their inherent metastability can also be exploited to design novel architectures such as those in galvanic replacements [16-20]. Dimer systems exploit proximity effects as source of extra functionality, but the contact interface area is limited thus precluding a large contact surface to promote interaction between the components. On the other hand, the encapsulation of core material into a core/shell structure promotes a large contact surface area, since the whole surface of the core becomes the interface with the shell. Additionally, it allows controlling the structure degree of freedom in the form of lattice constraints that permit modifying the system structure and properties. Interface related lattice constraints can result in thermodynamically metastable systems which may phase-segregate or homogenize due to ion diffusion upon heating, during electrochemical cyclic, or in long term ambient conditions [21], relevant to processes such as redox limited performance of energy materials. In this context, the study of processes that mimic the destabilization of hybrid core/shell nanoparticles under controlled conditions can facilitate the understanding of their behavior in real applications (e.g., in the study of cathode materials for batteries or electrochemical capacitors) and open new avenues to create improved structures [22-24].

This is particularly the case with nanostructured systems based on complex oxides, which are very robust materials displaying a wide plethora of macroscopic functionalities which are extremely sensitive to the oxygen stoichiometry. Minuscule changes in the local O content may result in large macroscopic responses, both due to variations in local electronic properties and also in lattice strains. In this context, investigations of the in-situ reduction of oxide nanoparticles provide a unique tool to understand the transformations that take place when heated [25] or upon redox reactions [16,17,26], likely scenarios in many of their applications [27-30]. Unfortunately, more often than not these processes take place in a very inhomogeneous fashion within nanometric length scales. Thus, monitoring nanometric reaction fronts in real space with atomic resolution becomes a key task which may pave the way to harnessing novel heterostructured nanosystems [31-34]. Macroscopically-averaged diffraction techniques cannot always provide a clear picture of the intrinsically discontinuous nature of such systems, hindering our understanding of the mechanisms underlying the behavior of such small active regions. Real space probes capable of providing a comprehensive view of the structure, chemistry and electronic properties in real space constitute the keystone towards harnessing hybrid nanosystems. Owing to the success of the correction of optical aberrations, advanced electron microscopy and spectroscopy techniques have undergone a major revolution within the last decade and can provide atomically resolved views of complex materials at the sub-Ångström scale with single-atom sensitivity [35-37]. Armed with such tools, we are in a unique position to address the study of heterogeneous nanostructured materials in real-time.

In this work, we present an atomic resolution live (in-situ) study of electron beam-induced transformations in core/shell $Fe_3O_4/Mn_3O_4$ and $Mn_3O_4/Fe_3O_4$ nanoparticles by means of high-resolution scanning transmission electron microscopy combined with electron energy-loss spectroscopy (STEM-EELS). Interestingly, the manganese oxide-iron oxide nanoparticles system is appealing not only for its unique magnetic properties (e.g., antiferromagnetic interface coupling or magnetic proximity effects) [10,38-40] but also for its applications in diverse fields like catalysis and environmental remediation [41,42], batteries and supercapacitors [43,44], microwave components [45] or biomedicine [46,47]. Our results, acquired within time scale of minutes, show that the nanoparticles under electron irradiation can develop a variety of defects, including formation of voids and dislocations, as the $Mn_3O_4$ component is forced to reduce to MnO. Such transformations depend on factors such as the morphology and they occur in different stages that include both redox reactions and generation of voids followed by stress relaxation, ionic transport and formation of defects.

## 2. Matrials and Methods

### 2.1 Synthesis of the $Fe_3O_4$/$Mn_3O_4$ and $Mn_3O_4$/$Fe_3O_4$ nanoparticles

$Fe_3O_4$/$Mn_3O_4$ and $Mn_3O_4$/$Fe_3O_4$ core/shell nanoparticles were synthesized using a seeded-growth approach [10], where pre-synthesized Fe-oxide and Mn-oxide nanoparticles were used as seeds (cores) for the growth of Mn-oxide and Fe-oxide shells, respectively. $Fe_3O_4$/$Mn_3O_4$ nanoparticles were synthesized using a slurry of dispersed iron oxide cores in dibenzyl ether heated up to 220 ºC. 10 mL of a dibenzyl ether solution containing manganese (II) acetylacetonate, 1,2-hexadecanediol, and a mixture of oleylamine and oleic acid as surfactants was injected rapidly to the slurry. The temperature was maintained at 220 ºC for 1 hour after the injection under magnetic stirring. Alternatively, $Mn_3O_4$/$Fe_3O_4$ nanoparticles were synthesized by dispersing manganese oxide cores in a dibenzyl ether solution of iron (III) acetylacetonate, 1,2-hexanedcandiol, oleic acid and oleylamine. The reaction flask was heated to 250 ºC for 20 minutes. The reaction vessel was removed from the heat and allowed to cool down to room temperature under argon. Both types of nanoparticles were washed by several cycles of coagulation with ethanol, centrifuging at 2000 g, disposal of supernatant, and redispersion in n-hexane.

**2.2 Scanning Transmission Electron Microscopy (STEM)**

STEM samples were prepared by sonicating the nanoparticles in methanol to make a suspended solution. One drop of the solution was deposited on a holey carbon TEM grid, and then the grid was air-dried. The experiments were performed on a Nion UltraSTEM200, equipped with a cold field-emission electron source and a fifth -order aberration corrector, operating at accelerating voltages of 60 kV and 200 kV. High angle annular dark field (HAADF) images were used to measure the lattice parameters of the different compounds. For this aim, a script was used to assign atomic column positions to the brightest pixel locations. Once all the coordinates of all atomic columns were obtained, distances between nearest neighbors were measured between them.

**2.3 Electron energy-loss spectra (EELS)**

Electron energy-loss spectra (EELS) were collected using a Gatan Enfinium spectrometer, with an energy dispersion of 0.25 eV per channel; the convergence semi-angle for the incident probe was 30 mrad, with an EELS collection semi-angle of 36 mrad. Electron beam currents were kept in the range of a few tens of pA, in a relatively high dose range estimated to be near $5\times10^{-7}$ e$^-$/Å$^2$. EEL spectrum images were acquired for the different elements of interest. Mn, Fe and O maps were produced using the Mn $L_{2,3}$, Fe $L_{2,3}$, and O $K$ edges, respectively. Spectrum images were acquired on nanoparticles that were off-axis in order to prevent channeling related artifacts during quantification. In order to carry out a meaningful quantification the background was subtracted using a power-law fit and the intensity under every absorption edge was integrated using a 25 eV wide window. We used the built-in quantification process available in Digital Micrograph (Gatan Inc) using the available cross sections. Principal component analysis (PCA) [48] was used to improve the signal-to-noise level in the spectrum images. Multiple linear least square fits were also used to analyze the spatial distribution of the different components (see Fig. S1 in the Electronic Supplementary Material (ESM) for an example). Note that at least three different on-axis, randomly selected particles were analyzed in each sample by means of atomic resolution STEM-EELS (previously, low magnification TEM characterization was carried out to ensure sample homogeneity).

## 2.4 STEM-HAADF simulations

STEM-HAADF simulations were obtained by a multi-slice method using the JEMS software. The experimental parameters used for the simulation are $C_S$ = 0.009 mm (spherical aberration) and $C_5$ = 10 mm (fifth-order spherical aberration), and all other axial aberration coefficients were neglected.

## 2.5 Geometric phase analysis (GPA)

Geometric phase analysis (GPA) was used to measure the distortion of lattice fringes with respect to a reference region in an HRTEM image of the particles [49]. A careful analysis was carried out to rule out the influence of scanning artifacts on the analysis. The lattice parameters were determined from the STEM images by averaging the distance between consecutive atoms in a given atom row (see example in Fig. S2 in the ESM). Geometric phase analysis was performed using 220 MnO/ 440 $Fe_3O_4$ and 002 MnO/ 004 $Fe_3O_4$ reflections using a cosine mask with a spatial resolution of 0.9 nm. The reference was chosen in the $Fe_3O_4$ unstrained region (top left region of the particle).

## 2.6 Particle image velocimetry (PIV)

The analyses of the STEM-HAADF images by particle image velocimetry (PIV) were carried out using the PIV plugin [50] for Image J [51]. Prior to the analysis, one hundred STEM-HAADF images (recorded every 2 seconds) were aligned to correct the drift, and a set of twenty images, obtained by summing up every five subsequent STEM-HAADF images, were used for the PIV analysis.

## 3. Results and Discussion

The studied core/shell $Fe_3O_4/Mn_3O_4$ and $Mn_3O_4/Fe_3O_4$ particles are highly crystalline and due to the topotactical growth of the spinel structures they exhibit defect free core/shell interfaces, as depicted in Fig. 1a and 1b. The morphology of the two systems is slightly different, where $Fe_3O_4/Mn_3O_4$ nanoparticles present an incomplete shell while $Mn_3O_4/Fe_3O_4$ nanoparticles exhibit a continuous, complete shell. The EELS maps shown in Fig. 1 reveal that the $Fe_3O_4/Mn_3O_4$ nanoparticles indeed consist of an iron oxide core with an overgrown manganese-rich shell (Fig. 1c-d and g-h). The stoichiometry of the $Fe_3O_4/Mn_3O_4$ core/shell nanoparticles was assessed by EELS off-axis imaging [52] both at 60 kV (Fig. 1e-f) and 200 kV (Fig. 1i-j). The false-color map displaying the relative oxygen content [53] of the $Fe_3O_4/Mn_3O_4$ core/shell (Fig. 1e) obtained at 60 kV reveals a fairly homogeneous stoichiometry across the nanoparticle. The laterally averaged line scan profile (Fig. 1f, where O stands for oxygen and M for the metal species) indicates relative oxygen to metal content of approximately 57 at%, in good agreement with the expected stoichiometry of the $Fe_3O_4$ and $Mn_3O_4$ phases. On the other hand, the analysis of equivalent data acquired at 200 kV (bottom panels) exhibits a severe inhomogeneity of the relative O content revealed through a significant O reduction near the surfaces (Fig. 1i). The laterally averaged profile (Fig. 1j) shows that the oxygen content decreases from ca. 57-59 at% near the central portion of the particle down to 50 at% at the surface of the $Fe_3O_4/Mn_3O_4$ core/shell, suggesting that the stoichiometry of the shell is, in fact, closer to that of MnO. We conclude that at 200 kV an electron beam-induced reduction of $Mn_3O_4$ to MnO takes place, in agreement with earlier reports [54,55]. Interestingly, this beam-induced phenomenon also takes place in the case of the $Mn_3O_4/Fe_3O_4$ configuration (note the presence of MnO at the core of the nanoparticle in Fig. S1 in the ESM), when a $Mn_3O_4$ core is completely surrounded by $Fe_3O_4$ and no free Mn oxide surface is available, indicating that this is not a surface-related effect. Typically, the reduction of the spinel phase into the rock salt phase is observed at relatively high temperature [56] and pressure but it can also be triggered by an induced reduction process [57]. Note that no electron-induced reduction of Fe (III) to Fe (II) is observed in neither the $Mn_3O_4/Fe_3O_4$ core/shell nor the $Fe_3O_4/Mn_3O_4$ core/shell configurations. This finding can be understood both in terms of the difference in Gibbs free enthalpy between the $Mn_3O_4 \rightarrow MnO$ and $Fe_3O_4 \rightarrow FeO$ transformation [54] and the instability of $Fe_{1-x}O$ [58] (see ESM).

The local processes underlying the reduction transformation can be monitored via atomic resolution imaging. High-resolution STEM-high angle annular dark field (STEM-HAADF) images recorded along the spinel [1-10] zone axis allow tracking in-situ electron beam-induced selective reduction of $Mn_3O_4$ in both $Fe_3O_4/Mn_3O_4$ and $Mn_3O_4/Fe_3O_4$ core/shell systems. As summarized in Fig. 2a-d (detail in e and f), a contrast change is observed in a temporal series of consecutive images, which is associated with the formation of MnO upon electron exposure. Both core/shell systems display a high crystallinity and more coherent and defect free interfaces before the transformation of the spinel-type $Mn_3O_4$ to a rock salt-type MnO (see Fig. 2a and c; detail in e-f). Comparing Fig. 2a and 2c allows identifying the location of the manganese oxide (right outer most edge of the particle) and iron oxide regions.

Similarly, it is possible to observe the location of the manganese oxide in the center of the particle shown in Fig. 2c-d. All of the involved phases are based on an FCC oxygen sublattice that is preserved upon reduction, permitting a high lattice coherency.

An in-depth analysis of the images of the $Fe_3O_4/Mn_3O_4$ nanoparticle, shown in Fig. 2a, evidences that the lattice parameter of the cubic spinel (Fd$\bar{3}$m) $Fe_3O_4$ core lies around $a(Fe_3O_4)_{core}$ = 8.4 Å, i.e. similar to the bulk [58] value, which gradually increases towards the nanoparticle surface. The shell shows a change of lattice parameter, in which $\sqrt{2}a(Mn_3O_4)_{shell}$ = 8.5 Å and $c(Mn_3O_4)_{shell}$ = 9.5 Å, denoting an expansion of $\Delta a(Mn_3O_4)_{shell} \approx$ +4% (compared to the bulk [58] value, for the tetragonal spinel (I4$_1$/amd) $Mn_3O_4$, $a$ = 5.76 Å and $c$ = 9.44 Å) due to the interface with the $Fe_3O_4$ core (Figs. S2 and S3 in the ESM). In addition, upon the reduction of the manganese oxide shell, an overall decrease around ~2% of the nanoparticle size is observed, which should be mostly related to oxygen removal. The lattice parameters obtained for the Mn oxide shell after reduction (MnO; rock salt, Fm$\bar{3}$m) are $a(MnO)_{shell}$ = 4.6 Å, about 4% larger than the bulk.

In the case of the $Mn_3O_4/Fe_3O_4$ core/shell nanoparticle, before the transformation, the analysis of the image shown in Fig. 2c indicates that the lattice parameters of $Mn_3O_4$ and $Fe_3O_4$ lie around $\sqrt{2}a(Mn_3O_4)_{core}$ = 8.5 Å and $c(Mn_3O_4)_{core}$ = 9.0 Å and $a(Fe_3O_4)_{shell}$ = 8.5 Å, i.e., $\Delta a(Mn_3O_4)_{core} \approx$ +4% (expansion) and $\Delta c(Mn_3O_4)_{core} \approx$ –5% (compression) and $\Delta a(Fe_3O_4)_{shell} \approx$ +1%, with respect to the bulk. In this case, the smaller relative changes in the lattice parameters of the shell reflect its large volume fraction, which dominates over the core. In general, such comparisons to the bulk values show that both the $Mn_3O_4$ and $Fe_3O_4$ contract and/or expand to keep the system energy low by generating a coherent interface [2,60,61] in which the configuration, i.e., $Mn_3O_4$ in the core or in the shell, and the volume fractions of each phase are very important to understand which phase undergoes the largest deformation. In fact, for $Fe_3O_4/Mn_3O_4$, with an incomplete shell, only the shell suffers strains to accommodate the coherent interface (see Fig. S4 in the ESM), whereas for $Mn_3O_4/Fe_3O_4$, with a fully developed core/shell structure, both counterparts need to strain their structures to stay structurally coherent. This indicates that the amount of elastic energy contained in the particles is still lower than that required to break the interface by forming an incoherent, defect rich state. Notably, these changes in their lattice parameters (i.e., strains) can ultimately lead to a large modification of their functional properties, e.g. their magnetic ground state, since the compressed (or expanded) lattice parameters effectively reduces (or enlarges) the interatomic distances. In the case of magnetic materials, a change of interatomic distances and/or angles directly influences the exchange coupling constant [62]. For example, for bulk MnO a reduction of the Mn-O bond length increases its Néel temperature [63,64], while strains can strongly influence the Verwey transition temperature of $Fe_3O_4$ [65,66].

The spinel and rock salt structures (e.g., $Mn_3O_4$ and MnO) transform topotactically [63] into one another via redistribution of cations from/to the interstitial tetrahedral positions to/from the octahedral positions while maintaining the oxide fcc-sublattice along with release/uptake of oxygen. The transformation observed in both core/shell systems (i.e. $Mn_3O_4 \rightarrow 3MnO + ½O_2$) involves a volume loss associated with oxygen loss and a lattice expansion/contraction (i.e., $\Delta a(Mn_3O_4) \rightarrow MnO = +8\%$ and $\Delta c(Mn_3O_4) \rightarrow MnO = -6\%$ for bulk materials). The comparison of the electron beam-induced reduction of the $Mn_3O_4$ in the two systems (Fig. 2b and 2d) suggests that the underlying mechanism of reduction in the $Mn_3O_4/Fe_3O_4$ core/shell nanoparticles might be more complex than in the case of $Fe_3O_4/Mn_3O_4$. When the $Mn_3O_4$ composes the shell, its reduction towards MnO modifies the lattice parameters as the reduction of Mn(III) to Mn(II) removes the Jahn-Teller distortion and increases the cation radius [67-70]. The associated large changes of lattice parameters (i.e., $\Delta a$(Mn$_3$O$_4$)→MnO ≈ +12 % and $\Delta c$(Mn$_3$O$_4$)→MnO ≈ –3 %) are easily accommodated by the absence of confinement which allows lattice rotation that maintains a defect-free coherent interface (see Fig. 2a-b and Fig. S4 in the ESM). Moreover, the lattice expansion/contraction evolves as expected during the transformation (compare the expansion/contraction sign of the lattice parameters with bulk materials). Conversely, in the case of the Mn$_3$O$_4$/Fe$_3$O$_4$ core/shell system, the Mn$_3$O$_4$ reduction comprises more intricate effects such as the concomitant formation of a void as the Mn$_3$O$_4$ core transforms into MnO (white circle in Fig. 2d) and the lattice expansion along the $c$-axis. Fig. 2e and 2f show a zoomed-in region of the interface at the core-shell before and after reduction, respectively. Since some Mn/Fe interdiffusion can take place during the transformation, it was allowed within the simulations to include a possible chemical intermixing, although no major effect was found. A comparison of the real image with the simulations indicate that the interface is relatively sharp on the order of one unit cell. Considering that the reduction of Mn$_3$O$_4$ to MnO in the Mn$_3$O$_4$/Fe$_3$O$_4$ case is more complex than for the Fe$_3$O$_4$/Mn$_3$O$_4$ case, we will study in more depth the evolution of the Mn$_3$O$_4$/Fe$_3$O$_4$ nanoparticles.

In order to monitor the atomic column movement associated with the reaction front in the Mn$_3$O$_4$/Fe$_3$O$_4$ nanoparticles, we used particle image velocimetry (PIV, see Fig. S5 in the ESM) [71,72] in a series of subsequent high-resolution STEM-HAADF images at early transformation times. Remarkably, the PIV analysis (see Fig. 3, S5 and Video V1 in the ESM) reveals an inhomogeneous pattern of column displacements (for our analysis we are assuming that the small thickness of the particles does not constitute an issue). The larger shifts are mainly localized at the top-left region of the particle shell towards the nanoparticle core, with a maximum displacement value of 1.5 Å (see Fig.

S5 in the ESM). The relatively small displacements observed at the right bottom corner are probably due to the pinning of the particle by other neighboring nanoparticles. The column displacement towards the core occurs as a strain relaxation mechanism related to oxygen loss during reduction of $Mn_3O_4$ to $MnO$. According to the PIV analysis, some cation columns move around the network during the beam-induced transformation. However, confinement effects within the MnO core in the $Mn_3O_4/Fe_3O_4$ core/shell nanoparticle prevent stress relief mechanisms via lattice rotation from taking place, a mechanism which is available for the $Fe_3O_4/Mn_3O_4$ configuration (see Fig. S4 in the ESM). As a consequence, misfit dislocations appear, as evidenced in Fig. 4 (see below). Surprisingly, in the $Mn_3O_4/Fe_3O_4$ core/shell nanoparticle the reduction of the $Mn_3O_4$ spinel structure brings about a global expansion of the lattice parameters along the *c*-axis of the core, (i.e., $\Delta c$ $Mn_3O_4 \rightarrow MnO \approx +4$ %, see Fig. S2 in the ESM). The unexpected lattice expansion of the core along the *c*-axis also compensates for the oxygen loss to maintain the coherent interface between the core and the shell of the nanoparticle. The reduction of Mn(III) to Mn(II) also results in an expected expansion of the *a*-axis within the core of about $\Delta a$ ($Mn_3O_4 \rightarrow MnO$) $\approx +5$ %. The expansion of the *a*-axis in the core results in the expansion of the *a*-axis of the shell as well.

Geometric phase analysis (GPA) [49] can be used for further analysis of the strains and defects resulting from the loss of oxygen and redistribution of cationic columns. Fig. 4 shows a quantification of the strain along the *c*-axis and along the [110] direction. The GPA analysis, using Fourier filtering, highlights the presence of defects in the $Mn_3O_4/Fe_3O_4$ core/shell nanoparticle upon electron exposure. The magnified STEM-HAADF images of the $Mn_3O_4/Fe_3O_4$ core/shell interface reveal structural changes upon electron exposure (Fig. 4a-d). The coherent spinel domain transforms into a metastable MnO-like phase (Fig. 4b), followed by a cation redistribution across the core/shell interface. We believe that upon prolonged exposure time the formation of an intermediate $Mn_xFe_{3-x}O_4$ phase is possible (Fig. 4d and also the detailed interface in Fig. 2f) although a definitive conclusion in this respect cannot be drawn from our data. The formation of this intermediate phase would be driven by an uneven cation distribution over the whole core/shell nanoparticle combined with a vacancy gradient between the inner and outer boundaries [73]. In fact, the deformation map shown in Fig. 4e shows that the core of the nanoparticle at early stages of exposure is positively strained (3-12%) along the *c*-axis with respect to the $Fe_3O_4$ shell (the down-right region of the nanoparticle shell is used as reference area for GPA), pointing towards the heterogeneous distribution of Mn and Fe cations [69]. Note that an absence of observable dislocations (from this particular zone axis) at the core/shell interface is observed in Fig. 4e-h (note that dislocations would be noticeable as discontinuities, i.e., sudden color change from yellow to blue, on the deformation map). This fact indicates that the nanoparticle core/shell boundary acts as a strained buffer layer of intermediate metastable cation arrangement in between the normal and inverse spinel structure, as previously reported on similar γ-$Fe_2O_3$/$Mn_3O_4$ core/shell nanoparticles [74,75]. The deformation maps shown in Fig. 4f-h reveal a positively strained core along with the presence of interfacial defects. The dislocations act as a strain relief at the core/shell nanoparticle interface allowing the nanoparticle to maintain a stable interface since the confinement of the core does not allow lattice rotation. These defects are misfit dislocations that are characterized by a shift of ¼ along [001] with respect to the $Fe_3O_4$ lattice, or alternatively by a shift of ½ [001] with respect to the MnO lattice, as evidenced by white circles in panel j. The misfit dislocation pairs, in turn, propagate around the core/shell interface but do not annihilate (see Fig. 4k,l). The deformation maps along [110] indicate a relatively homogeneous strain throughout the nanoparticle with some small local interfacial strain (see Fig. 4m-p). This finding indicates that the $Mn_3O_4/Fe_3O_4$ core/shell nanoparticle is anisotropically strained, so possible strain relaxation mechanisms involving cation redistribution along the *c*-axis towards the interface can be expected upon annealing (external or under the electron beam). Also, the core of the nanoparticle remains in a relatively highly strained state (3-12%) as the volume reduction associated with oxygen loss contributes to increasing the strain at the nanoparticle interface. Defects such as misfit dislocations are observed (Fig. 4q-t). The formation of MnO implies a large displacement of cations from tetrahedral into octahedral positions. This fact, together with the presence of newly generated Mn(II) ions have been reported in similar cases before, is again likely to promote the formation of interfacial $Mn_xFe_{3-x}O_4$ [3,74,75].

## 4 Conclusion

In summary, we have demonstrated that a selective electron-induced structural transformation of $Mn_3O_4$ into MnO in $Fe_3O_4/Mn_3O_4$ and $Mn_3O_4/Fe_3O_4$ core/shell nanoparticle systems can take place through different mechanisms depending on the core/shell configuration. The confinement of the $Mn_3O_4$ in $Mn_3O_4/Fe_3O_4$ core/shell nanoparticle prevents stress relief by lattice rotation and results in the formation of misfit dislocation pairs. Prolonged exposure of the particles, may ultimately lead to annealing of the core/shell nanoparticles leading to an equilibrium single phase. We have shown how in-situ STEM-HAADF monitoring of the atomic column positions and image analysis combining GPA and PIV are useful tools to study chemical transformations and diffusion processes in solid materials as they allow direct observation of the relationship between the structure and strain relaxation during the $Mn_3O_4$ reduction. Such electron beam-induced transformations can be tuned by varying the acceleration voltage, which can be exploited to further modulate the energy transfer in a controlled and reproducible manner. The selective reduction of confined heterostructured materials based on functional oxides (e.g., core/shell nanoparticles) systems offers innovative possibilities for locally tuning or patterning their electronic, magnetic and other functional properties (e.g., $Mn_3O_4$ is ferrimagnetic while MnO is antiferromagnetic) in order to optimize functionality in fields as diverse as catalysis, magnetism, biomedicine, or sensors.

**Declaration of Competing Interest**

The authors declare that they have no known competing financial interests or personal relationships that could have appeared to influence the work reported in this paper.

**Acknowledgements**

Research supported by the European Research Council Starting Investigator Award STEMOX # 239739 (M.R. and J. S.), JSPS Postdoctoral Fellowship for Research Abroad (R.I.) and by Spanish MAT2015-066888-C3-3-R and RTI2018-097895-B-C43 (MINECO/FEDER). Electron microscopy observations at ORNL supported by the U.S. Department of Energy (DOE), Basic Energy Sciences (BES), Materials Sciences and Engineering Division and through a user project supported by ORNL's Center for Nanophase Materials Sciences (CNMS), which is sponsored by the Scientific User Facilities Division, Office of Basic Energy Sciences, U.S. Department of Energy. A.M. and G.S.A. thank the financial support of the Knut and Alice Wallenberg Foundation through the project 3DEM-NATUR. The work at INC2 has been supported by the 2017-SGR-292 project of the Generalitat de Catalunya and by the MAT2016-77391-R project of the Spanish MINECO. ALO acknowledges the Spanish Ministerio de Economía y Competitividad through the Juan de la Cierva Program (IJCI-2014-21530). ICN2 is funded by the CERCA Programme/Generalitat de Catalunya. ICN2 also acknowledges support from the Severo Ochoa Centres of Excellence programme, funded by the Spanish Research Agency (AEI, grant no. SEV-2017-0706). S. J. P. thanks the National University of Singapore for funding. M. E. Acknowledges the Spanish MINECO for her Ramón y Cajal Fellowship (RYC2018-024396-I).

**Appendix A. Supplementary data**

Supplementary data to this article can be found online at:

**References**

[1] P. D. Cozzoli, T. Pellegrino, L. Manna, Synthesis, properties and perspectives of hybrid nanocrystal structures. Chem. Soc. Rev. 35 (2006) 1195-1208, https://doi.org/10.1039/b517790c

[2] N. J. O. Silva, M. Karmaoui, V. S. Amaral, I. Puente-Orench, J. Campo, I. da Silva, A. Ibarra, R. Bustamante, A. Millán, F. Palacio, Shell pressure on the core of MnO/$Mn_3O_4$ core/shell nanoparticles. Phys. Rev. B 87 (2013) 224429, https://doi.org/10.1103/PhysRevB.87.224429

[3] K. L. Krycka, J. A. Borchers, G. Salazar-Alvarez, A. López-Ortega, M. Estrader, S. Estradé, E. Winkler, R. D. Zysler, J. Sort, F. Peiró, M. D. Baró, C.- C. Kao, J. Nogués, Resolving material-specific structures within $Fe_3O_4$|γ-$Mn_2O_3$ core|shell nanoparticles using anomalous small-angle x‑ray scattering. ACS Nano 7 (2013) 921-931, https://doi.org/10.1021/nn303600e

[4] W. Shi, H. Zeng, Y. Sahoo, T. Y. Ohulchanskyy, Y. Ding, Z. L. Wang, M. Swihart, P. N. Prasad, A general approach to binary and ternary hybrid nanocrystals. Nano Lett. 6 (2006) 875-881, https://doi.org/10.1021/nl0600833

[5] G. Salazar-Alvarez, H. Lidbaum, A. López-Ortega, M. Estrader, K. Leifer, J. Sort, S. Suriñach, M. D. Baró, Nogués, Two-, three-, and four-component magnetic multilayer onion nanoparticles based on iron oxides and manganese oxides. J. Am. Chem. Soc. 133 (2011) 16738-16741, https://doi.org/10.1021/ja205810t

[6] P. Torruella, A. Ruiz-Caridad, M. Walls, A. G. Roca, A. Lopez-Ortega, J. Blanco-Portals, L. Lopez-Conesa, J. Nogues, F. Peiro, S. Estrade, Atomic-scale determination of cation inversion in spinel-based oxide nanoparticles. Nano Lett. 18 (2018) 5854-5861, https://doi.org/10.1021/acs.nanolett.8b02524

[7] Y. Lee, M. A. Garcia, N. A. Frey Huls, S. Sun, Synthetic tuning of the catalytic properties of Au-$Fe_3O_4$ nanoparticles. Angew. Chemie Int. Ed. 49 (2010) 1271-

1274, https://doi.org/10.1002/anie.200906130

[8] P. Reiss, M. Protière, L. Li, Core/shell semiconductor nanocrystals. Small 5 (2009) 154-168, https://doi.org/10.1002/smll.200800841

[9] M. B. Cortie, A. M. McDonagh, Synthesis and optical properties of hybrid and alloy plasmonic nanoparticles. Chem. Rev. 111 (2011) 3713-3735, https://doi.org/10.1021/cr1002529

[10] M. Estrader, A. López-Ortega, S. Estradé, I. V. Golosovsky, G. Salazar-Alvarez, M. Vasilakaki, K. N. Trohidou, M. Varela, D. C. Stanley, M. Sinko, M. J. Pechan, D. J. Keavney, F. Peiró, S. Suriñach, M. D. Baró, J. Nogués, Robust antiferromagnetic coupling in hard-soft bi-magnetic core/shell nanoparticles Nat. Commun. 4 (2013) 2960, https://doi.org/10.1038/ncomms3960

[11] A. López-Ortega, M. Estrader, G. Salazar-Alvarez, A. G. Roca, J. Nogués, Applications of exchange coupled bi-magnetic hard/soft and soft/hard magnetic core/shell nanoparticles. Phys. Rep. 553 (2015) 1-32, https://doi.org/10.1016/j.physrep.2014.09.007

[12] Y. Jin, C. Jia, S.-W. Huang, M. O'Donnell, X. Gao, Multifunctional nanoparticles as coupled contrast agents. Nat. Commun. 1 (2010) 41, https://doi.org/10.1038/ncomms1042

[13] J.-H. Lee, J.-T. Jang, J.-S. Choi, S. H. Moon, S.-H. Noh, J.-W. Kim, J.-G. Kim, I.-S. Kim, K. I. Park, J. Cheon, Exchange-coupled magnetic nanoparticles for efficient heat induction. Nat. Nanotechnol. 6 (2011) 418-422, https://doi.org/10.1038/nnano.2011.95

[14] I. S. H. Lee, N. Lee, J. Park, B. H. Kim, Y.-W. Yi, T. K. Kim, S. R. Paik, T. Hyeon, Ni/NiO core/shell nanoparticles for selective binding and magnetic separation of histidine-tagged proteins. J. Am. Chem. Soc. 128 (2006) 10658-10659, https://doi.org/10.1021/ja063177n

[15] N.-H. Cho, T.-C. Cheong, J. H. Min, J. H. Wu, S. J. Lee, D. Kim, J.-S. Yang, S. Kim, S. Y. K. Kim, S.-Y. Seong, A multifunctional core–shell nanoparticle for dendritic cell-based cancer immunotherapy. Nat. Nanotechnol. 6 (2011) 675-682, https://doi.org/10.1038/nnano.2011.149

[16] M. H. Oh, T. Yu, S.-H. Yu, B. Lim, K.-T. Ko, M.-G. Willinger, D.-H. Seo, B. H. Kim, M. G. Cho, J.-H. Park, K. Kang, Y.-E. Sung, N. Pinna, T. Hyeon, Science 340 (2013) 964-968, https://doi.org/10.1126/science.1234751

[17] A. López-Ortega, A. G. Roca, P. Torruella, M. Petrecca, S. Estradé, F. Peiró, V. Puntes, J. Nogués, Galvanic replacement onto complex metal-oxide nanoparticles: Impact of water or other oxidizers in the formation of either fully dense onion-like or multicomponent hollow $MnO_x/FeO_x$ structures. Chem. Mater. 28 (2016) 8025-8031, https://doi.org/10.1021/acs.chemmater.6b03765

[18] T. Teranishi, Y. Inoue, M. Nakaya, Y. Oumi, T. Sano, Nanoacorns: anisotropically phase-segregated CoPd sulfide nanoparticles. J. Am. Chem. Soc. 126 (2004) 9914-9915, https://doi.org/10.1021/ja047606y

[19] J. Yang, J. Peng, Q. Zhang, F. Peng, H. Wang, H. Yu, One‑step synthesis and characterization of gold‑hollow $PbS_x$ hybrid nanoparticles. Angew. Chemie Int. Ed. 48 (2009) 3991-3995, https://doi.org/10.1002/anie.200806036

[20] A. A. Dzhurakhalov, M. Hou, Equilibrium properties of binary and ternary metallic immiscible nanoclusters. Phys. Rev. B 76 (2007) 045429, https://doi.org/10.1103/PhysRevB.76.045429

[21] L. Carbone, P. D. Cozzoli, Colloidal heterostructured nanocrystals: synthesis and growth mechanisms. Nano Today 5 (2010) 449-493, https://doi.org/10.1016/j.nantod.2010.08.006

[22] A.- A. El Mel, L. Molina-Luna, M. Buffière, P.- Y. Tessier, K. Du, C.- H. Choi, H.- J. Kleebe, S. Konstantinidis, C. Bittencourt, R. Snyders, Electron beam nanosculpting of kirkendall oxide nanochannels. ACS Nano 8 (2014) 1854-1861, https://doi.org/10.1021/nn406328f

[23] F. E. Şeşen, Practical reduction of manganese oxide. J. Chem. Tech. Appl. 1 (2017) 26-27, https://doi.org/10.35841/chemical-technology.1.1.26-27

[24] P. Iamprasertkun, A. Krittayavathananon, A. Seubsai, N. Chanlek, P. Kidkhunthod, W. Sangthong, S. Maensiri,, R. Yimnirun, S. Nilmoung, P. Pannopard, S. Ittisanronnachai, K. Kongpatpanich, J. Limtrakul, M. Sawangphruk, Charge storage mechanisms of manganese oxide nanosheets and N-doped reduced graphene oxide aerogel for high-performance asymmetric supercapacitors. Sci. Rep. 6 (2016) 37560, https://doi.org/10.1038/srep37560

[25] J. Pike, J. Hanson, L. Zhang, S. Chan, Synthesis and redox behavior of nanocrystalline hausmannite ($Mn_3O_4$). Chem. Mater. 19 (2007) 5609-5616, https://doi.org/10.1021/cm071704b

[26] J. Graetz, C. C. Ahn, H. Ouyang, P. Rez, B. Fultz, White lines and d-band occupancy for the 3d transition-metal oxides and lithium transition-metal oxides. Phys. Rev. B 69 (2004) 235103, https://doi.org/10.1103/PhysRevB.69.235103

[27] M. Lundberg, Model calculations on some feasible two-step water splitting processes. Int. J. Hydrogen Energy 18 (1993) 369-376, https://doi.org/10.1016/0360-3199(93)90214-U

[28] J.- M. Tarascon, P. Poizot, S. Laruelle, S. Grugeon, L. Dupont, Nano-sized transition-metal oxides as negative-electrode materials for lithium-ion batteries. Nature 407 (2000) 496-499, https://doi.org/10.1038/35035045

[29] A. Dreyer, A. Feld, A. Kornowski, E. D. Yilmaz, H. Noei, A. Meyer, T. Krekeler, C. Jiao, A. Stierle, V. Abetz, H. Weller, G. A. Schneider, Organically linked iron oxide nanoparticle supercrystals with exceptional isotropic mechanical properties Nat. Mater. 15 (2016) 522-528, https://doi.org/10.1038/nmat4553

[30] Z. Sun, Y. Zhang, Y. Liu, J. Fu, S. Cheng, P. Cui, E. Xie, New insight on the mechanism of electrochemical cycling effects in $MnO_2$-based aqueous supercapacitor. J. Power Sources 436 (2019) 226795, https://doi.org/10.1016/j.jpowsour.2019.226795

[31] H. Yoshida, S. Takeda, T. Uchiyama, H. Kohno, Y. Homma, Atomic-scale in-situ observation of carbon nanotube growth from solid state iron carbide nanoparticles. Nano Lett. 8 (2008) 2082-2086, https://doi.org/10.1021/nl080452q

[32] F. Wang, H. -C. Yu, M. -H. Chen, L. Wu, N. Pereira, K. Thornton, A. Van der Ven, Y. Zhu, G. G. Amatucci, J. Graetz, Tracking lithium transport and electrochemical reactions in nanoparticles. Nat. Commun. 3 (2012) 1201, https://doi.org/10.1038/ncomms2185

[33] C.- M. Wang, A. Genc, H. Cheng, L. Pullan, D. R. Baer, S. M. Bruemmer, In-Situ TEM visualization of vacancy injection and chemical partition during oxidation of Ni-Cr nanoparticles. Sci. Rep. 4 (2014) 3683, https://doi.org/10.1038/srep03683

[34] C. G. Read, T. R. Gordon, J. M. Hodges, R. E. Schaak, Colloidal hybrid nanoparticle insertion reaction for transforming heterodimers into heterotrimers. J. Am. Chem. Soc. 137 (2015) 12514-12517, https://doi.org/10.1021/jacs.5b08850

[35] N. D. Browning, M. F. Chisholm, S. J. Pennycook, Atomic-resolution chemical analysis using a scanning transmission electron microscope. Nature 366 (1993) 143-146, https://doi.org/10.1038/366143a0

[36] P. D. Nellist, M. F. Chisholm, N. Dellby, O. L. Krivanek, M. F. Murfitt, Z. S. Szilagyi, A. R. Lupini, A. Borisevich, W. H. Sides, S. J. Pennycook, Direct sub-angstrom imaging of a crystal lattice. Science 305 (2004) 1741, https://doi.org/10.1126/science.1100965

[37] M. Varela, S. D. Findlay, A. R. Lupini, H. M. Christen, A. Y. Borisevich, N. Dellby, O. L. Krivanek, P. D. Nellist, M. P. Oxley, L. J. Allen, S. J. Pennycook, Spectroscopic imaging of single atoms within a bulk solid. Phys. Rev. Lett. 92 (2004) 095502, https://doi.org/10.1103/PhysRevLett.92.095502

[38] M. Jiang, X. Peng, Anisotropic $Fe_3O_4/Mn_3O_4$ hybrid nanocrystals with unique magnetic properties. Nano Lett. 17 (2017) 3570-3575, https://doi.org/10.1021/acs.nanolett.7b00727

[39] P. K. Manna, S. M. Yusuf, M. Basu, T. Pal, The magnetic proximity effect in a ferrimagnetic $Fe_3O_4$ core/ferrimagnetic $\gamma$-$Mn_2O_3$ shell nanoparticle system. J. Phys.: Condens. Matter 23 (2011) 506004, https://doi.org/10.1088/0953-8984/23/50/506004

[40] K. L. Krycka, J. A. Borchers, M. Laver, G. Salazar-Alvarez, A. López-Ortega, M. Estrader, S. Suriñach, M. D. Baró, J. Sort, J. Nogués, Correlating material-specific layers and magnetic distributions within onion-like $Fe_3O_4$/MnO/$\gamma$-$Mn_2O_3$ core/shell nanoparticles. J. Appl. Phys. 113 (2013) 17B531, https://doi.org/10.1063/1.4801423

[41] Z. Wan, J. Wang, Degradation of sulfamethazine using $Fe_3O_4$-$Mn_3O_4$/reduced graphene oxide hybrid as Fenton-like catalyst. J. Hazard. Mater. 324 (2017) 653-664, https://doi.org/10.1016/j.jhazmat.2016.11.039

[42] M. Ma, Y. Yang, Y. Chen, F. Wu, W. Li, P. Lyu, Y. Ma, W. Tan, W. Huang,

Synthesis of hollow flower-like $Fe_3O_4/MnO_2/Mn_3O_4$ magnetically separable microspheres with valence heterostructure for dye degradation. Catalysts 9 (2019) 589, https://doi.org/10.3390/catal9070589

[43] M. Wang, Y. Huang, Y. Zhu, M. Yu, X. Qin, H. Zhang, Core-shell $Mn_3O_4$ nanorods with porous $Fe_2O_3$ layer supported on graphene conductive nanosheets for high-performance lithium storage application. Composites Part B 167 (2019) 668-675, https://doi.org/10.1016/j.compositesb.2019.03.037

[44] R. Kumara, S. M. Youssry, K. Z. Ya, W. K. Tan, G. Kawamura, A. Matsuda, Microwave-assisted synthesis of $Mn_3O_4$-$Fe_2O_3$/$Fe_3O_4$@rGO ternary hybrids and electrochemical performance for supercapacitor electrode. Diamond Relat. Mater. 101 (2020) 107622, https://doi.org/10.1016/j.diamond.2019.107622

[45] R. B. Yang, W. F. Liang, C. K. Lin, Electromagnetic characteristics of manganese oxide-coated $Fe_3O_4$ nanoparticles at 2-18 GHz. J. Appl. Phys. 109 (2011) 07D722, https://doi.org/10.1063/1.3545810

[46] M. H. Kim, H. Y. Son, G. Y. Kim, K. Park, Y. M. Huh, S. Haam, Redoxable heteronanocrystals functioning magnetic relaxation switch for activatable $T_1$ and $T_2$ dual-mode magnetic resonance imaging. Biomaterials 101 (2016) 121-130, https://doi.org/10.1016/j.biomaterials.2016.05.054

[47] H. Rahaman, A. Nath, R. Paul, M. Sengupta, S. K. Ghosh, $Fe_3O_4$–$Mn_3O_4$ nanocomposites with moderate magnetism for in vitro cytotoxicity studies on macrophages. RSC Adv. 6 (2016) 83146-83153, https://doi.org/10.1039/C6RA17493K

[48] M. Bosman, M. Watanabe, D. T. L. Alexander, V. J. Keast, Mapping chemical and bonding information using multivariate analysis of electron energy-loss spectrum images. Ultramicroscopy 106 (2006) 1024-1032, https://doi.org/10.1016/j.ultramic.2006.04.016

[49] M. J. Hÿtch, E. Snoeck, R. Kilaas, Quantitative measurement of displacement and strain fields from HREM micrographs. Ultramicroscopy 74 (1998) 131-146, https://doi.org/10.1016/S0304-3991(98)00035-7

[50] Q. Tseng, PIV (Particle Image Velocimetry) - ImageJ plugin. https://sites.google.com/site/qingzongtseng/piv (accessed Mar 7, 2016).

[51] C. A. Schneider, W. S. Rasband, K. W. Eliceiri, NIH Image to ImageJ: 25 years of image analysis. Nat. Methods 9 (2012) 671-675, https://doi.org/10.1038/nmeth.2089

[52] R. F. Egerton, Electron energy-loss spectroscopy in the TEM. Rep. Prog. Phys. 72 (2009) 016502, https://doi.org/10.1088/0034-4885/72/1/016502

[53] L. Yedra, E. Xuriguera, M. Estrader, A. López-Ortega, M. D. Baró, J. Nogués, M. A. Roldan, M. Varela, S. Estradé, F. Peiró, Oxide Wizard: An EELS application to characterize the white lines of transition metal edges. Microsc. Microanal. 20 (2014) 698-705, https://doi.org/10.1017/S1431927614000440

[54] L. A. J. Garvie, A. J. Craven, Electron-beam-induced reduction of $Mn^{4+}$ in manganese oxides as revealed by parallel EELS. Ultramicroscopy 54 (1994) 83-92, https://doi.org/10.1016/0304-3991(94)90094-9

[55] S. J. Pennycook, L. Jones, H. Pettersson, J. Coelho, M. Canavan, B. Mendoza-Sanchez, V. Nicolosi, P. D. Nellist, Atomic scale dynamics of a solid state chemical reaction directly determined by annular dark-field electron microscopy. Sci. Rep. 4 (2014) 7555, https://doi.org/10.1038/srep07555

[56] J. Darul, C. Lathe, P. Piszora, $Mn_3O_4$ under high pressure and temperature: Thermal stability, polymorphism, and elastic properties. J. Phys. Chem. C 117 (2013) 23487-23494, https://doi.org/10.1021/jp404852j

[57] B. J. Kooi, J. T. M. De Hosson, In-situ TEM analysis of the reduction of nanometre-sized $Mn_3O_4$ precipitates in a metal matrix. Acta Mater. 49 (2001) 765-774, https://doi.org/10.1016/S1359-6454(00)00386-4

[58] M. W. Chase, NIST-JANAF thermochemical tables. Fourth Edition. J. Phys. Chem. Ref. Data Monographs 9 (1998) 1-1951.

[59] A. Navrotsky, C. Ma, K. Lilova, N. Birkner, Nanophase transition metal oxides show large thermodynamically driven shifts in oxidation-reduction equilibria. Science 330 (2010) 199-201, https://doi.org/10.1126/science.1195875

[60] G. Salazar-Alvarez, J. Sort, S. Suriñach, M. D. Baró, J. Nogués, Synthesis and size-dependent exchange bias in inverted core-shell $MnO|Mn_3O_4$ nanoparticles. J. Am. Chem. Soc. 129 (2007) 9102-9108, https://doi.org/10.1021/ja0714282

[61] E. Wetterskog, C.- W. Tai, J. Grins, L. Bergström, G. Salazar-Alvarez, Anomalous magnetic properties of nanoparticles arising from defect structures: Topotaxial oxidation of $Fe_{1-x}O|Fe_{3-\delta}O4$ core|shell nanocubes to single-phase particles. ACS Nano 7 (2013) 7132-7144, https://doi.org/10.1021/nn402487q

[62] S. Chikazumi, Physics of Magnetism, 2nd ed.; Oxford University Press, New York, 1997.

[63] D. Bloch, C. Vettier, P. Burlet, Phase transition in manganese oxide at high pressure. Phys. Lett. A 75 (1980) 301-303, https://doi.org/10.1016/0375-9601(80)90570-8

[64] S. Tamura, Pressure dependence of the Néel temperature of $MnO_{1+y}$ measured with a strain gauge to 2 GPa. High Temp. High Press. 19 (1987) 657-659.

[65] G. K. Rozenberg, G. R. Hearne, M. P. Pasternak, P. A. Metcalf, J. M. Honig, Nature of the Verwey transition in magnetite ($Fe_3O_4$) to pressures of 16 GPa. Phys. Rev. B 53 (1996) 6482-6487, https://doi.org/10.1103/PhysRevB.53.6482

[66] M. Estrader, A. López-Ortega, I. V. Golosovsky, S. Estradé, A. G. Roca, G. Salazar-Alvarez, L. López-Conesa, D. Tobia, E. Winkler, J. D. Ardisson, W. Macedo, A. Morphis, M. Vasilakaki, K. N. Trohidou, A. Gukasov, I Mirebeau, O. L. Makarova, R. D. Zysler, F. Peiró, M. D. Baró, L. Bergström, J. Nogués, Origin of the large dispersion of magnetic properties in nanostructured oxides:

Fe$_x$O/Fe$_3$O$_4$ nanoparticles as a case study. Nanoscale 7 (2015) 3002-3015, https://doi.org/10.1039/C4NR06351A

[67] R. D. Shannon, R. C. Rossi, Definition of topotaxy. Nature 202 (1964) 1000-1001, https://doi.org/10.1038/2021000a0

[68] H. F. McMurdie, B. M. Sullivan, F. A. Mauer, High Temperature X-ray Study of the System Fe$_3$O$_4$-Mn$_3$O$_4$. J. Res. Natl. Bur. Stand. 45 (1950) 35-41, https://doi.org/10.6028/jres.045.004

[69] P. Wojtowicz, Theoretical model for tetragonal-to-cubic phase transformations in transition metal spinels. Phys. Rev. 116 (1959) 32-45, https://doi.org/10.1103/PhysRev.116.32

[70] J. B. Goodenough, Jahn-Teller phenomena in solids. Annu. Rev. Mater. Sci. 28 (1998) 1-27, https://doi.org/10.1146/annurev.matsci.28.1.1

[71] R.J. Adrian, Twenty years of particle image velocimetry, Exp. Fluids 39 (2005) 159-169, https://doi.org/10.1007/s00348-005-0991-7

[72] E. R. White, M. Mecklenburg, B. Shevitski, S. B. Singer, B. C. Regan, Charged nanoparticle dynamics in water induced by scanning transmission electron microscopy. Langmuir 28 (2012) 3695-3698, https://doi.org/10.1021/la2048486

[73] A. M. Gusak, K. N. Tu, Interaction between the Kirkendall effect and the inverse Kirkendall effect in nanoscale particles. Acta Mater. 57 (2009) 3367-3373, https://doi.org/10.1016/j.actamat.2009.03.043

[74] A. Juhin, A. López-Ortega, M. Sikora, C. Carvallo, M. Estrader, S. Estradé, F. Peiró, M. D. Baró, P. Sainctavit, P. Glatzel, J. Nogués, Direct evidence for an interdiffused intermediate layer in bi-magnetic core–shell nanoparticles. Nanoscale 6 (2014) 11911-11920, https://doi.org/10.1039/C4NR02886D

[75] A. López-Ortega, M. Estrader, G. Salazar-Alvarez, S. Estradé, I. V. Golosovsky, R. K. Dumas, D. J. Keavney, M. Vasilakaki, K. N. Trohidou, J. Sort, F. Peiró, S. Suriñach, M. D. Baró, J. Nogués, Strongly exchange coupled inverse ferrimagnetic soft/hard, Mn$_x$Fe$_{3-x}$O$_4$/Fe$_x$Mn$_{3-x}$O$_4$, core/shell heterostructured nanoparticles. Nanoscale 4 (2012) 5138-5147, https://doi.org/10.1039/C2NR30986F

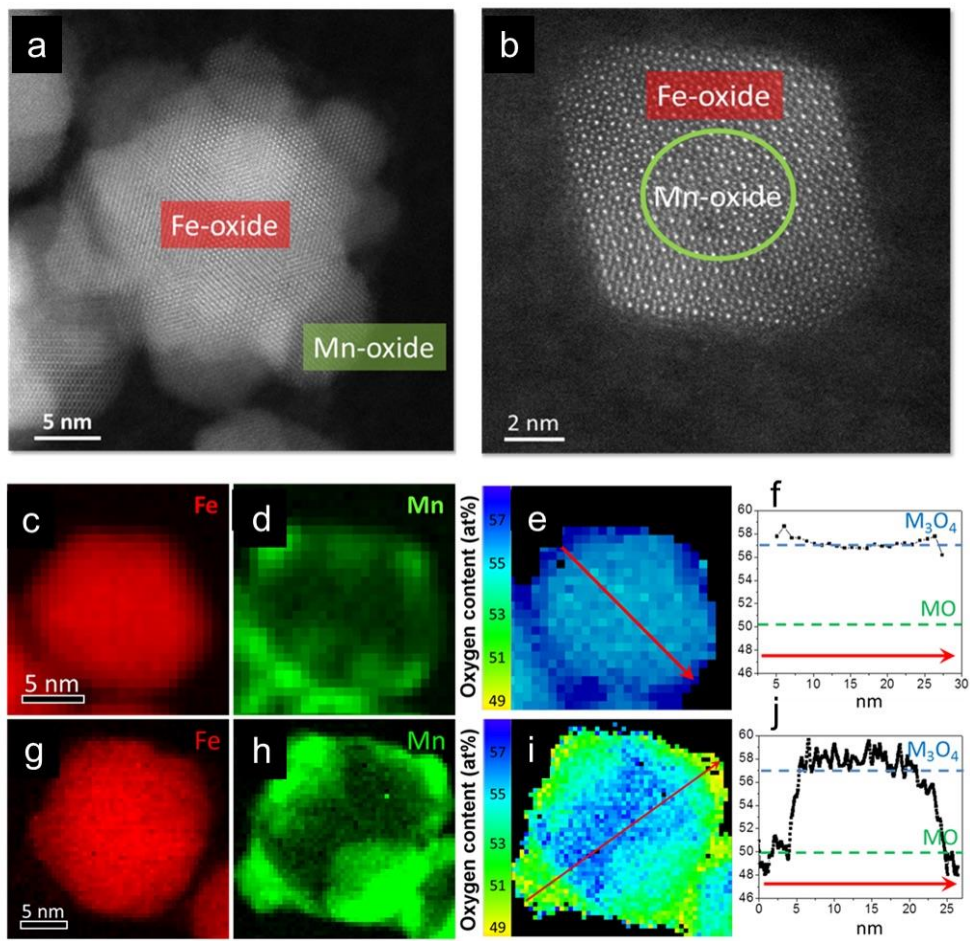

**Fig. 1.** High Resolution HAADF images of both types of core|shell nanoparticles. (a) $Fe_3O_4/Mn_3O_4$ and (b) $Mn_3O_4/Fe_3O_4$. Elemental mapping of two different $Fe_3O_4/Mn_3O_4$ core/shell nanoparticles at (c-f) 60 kV and (g-i) 200 kV. EELS maps obtained from the Fe $L_{2,3}$ edge at (c) 60 kV and (g) 200 kV. Mn $L_{2,3}$ edge at (d) 60 kV and (h) 200 kV. Maps of the relative oxygen content at (e) 60 kV and (i) 200 kV. Data along the direction outlined with a red arrow in (e) and (i), respectively, are shown in panels (f) and (j).

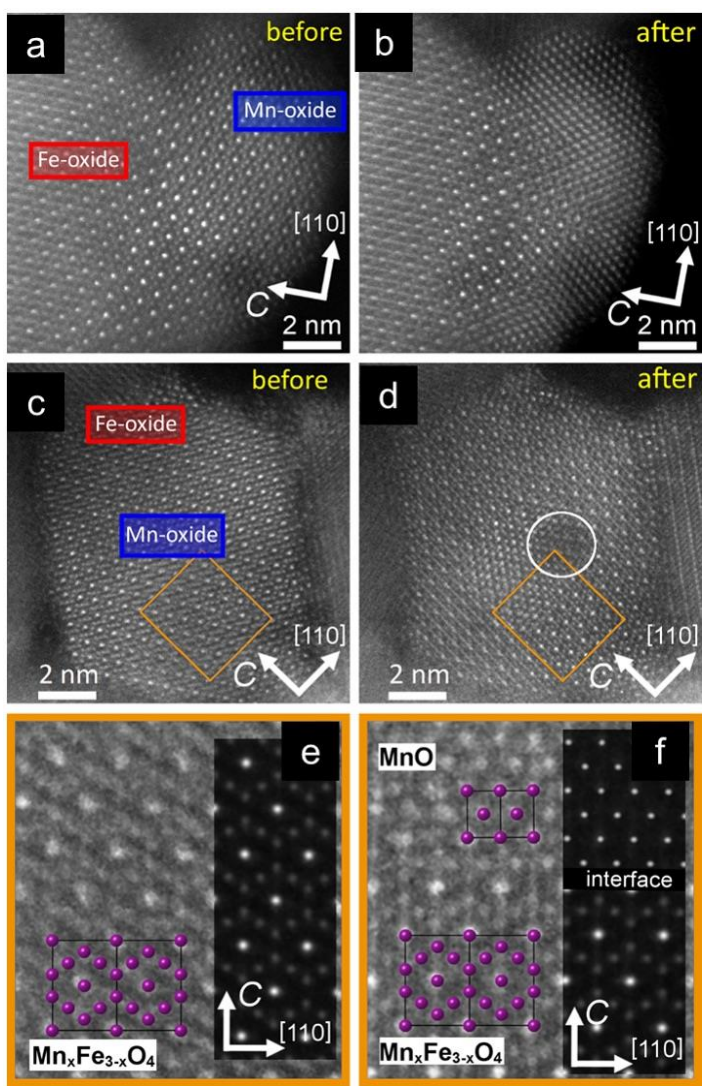

**Fig. 2.** Electron-induced transformation from $Mn_3O_4$ to MnO in $Fe_3O_4/Mn_3O_4$ and $Mn_3O_4/Fe_3O_4$ core/shell nanoparticles. (a-b) High-resolution STEM-HAADF images recorded along the [1-10] $Fe_3O_4$ axis of a $Fe_3O_4/Mn_3O_4$ core/shell nanoparticle (a) at an early stage and (b) after 10 minutes of electron beam exposure. (c-d) High-resolution STEM-HAADF images recorded along the [1-10] $Fe_3O_4$ axis of a $Mn_3O_4/Fe_3O_4$ core/shell nanoparticle (c) at an early stage and (d) after 30 minutes of beam exposure. The white circled area in (d) indicates the position of a void. (e-f) Magnified STEM-HAADF images of the highlighted (orange box) regions in (c) and (d), respectively, with the unit cells overlaid (oxide anions not shown). (Insets) Multislice simulation of a 12-nm thick $Mn_xFe_{3-x}O_4$ spinel and a MnO rock salt-type structures along [1-10].

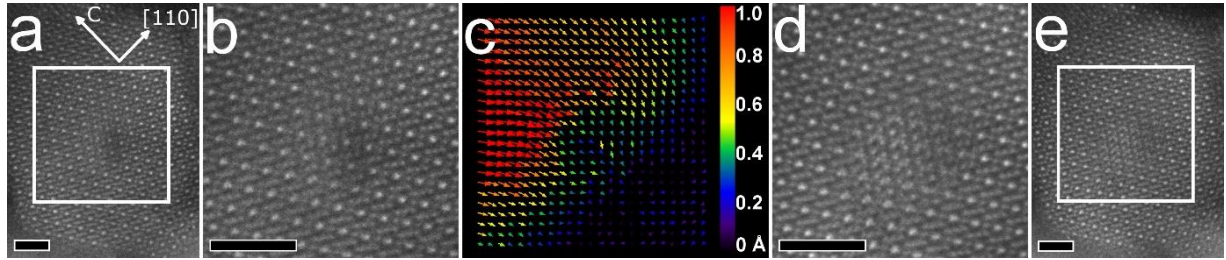

**Fig. 3.** Atomic column displacement analysis during phase transformation of the $Mn_3O_4$ core into MnO. (a) High-resolution STEM-HAADF image of the $Mn_3O_4/Fe_3O_4$ core/shell nanoparticle at the initial electron exposure. (b) Magnified STEM-HAADF image of the boxed area in image (a). (c) Displacement map obtained by PIV analysis of the STEM-HAADF images in (b) and (d). The false color scale indicates the atomic column displacement in Å. (d) Magnified STEM-HAADF image of the boxed area in image (e). (e) High-resolution STEM-HAADF image of the $Mn_3O_4/Fe_3O_4$ core/shell nanoparticle after 3.5 minutes of electron exposure.

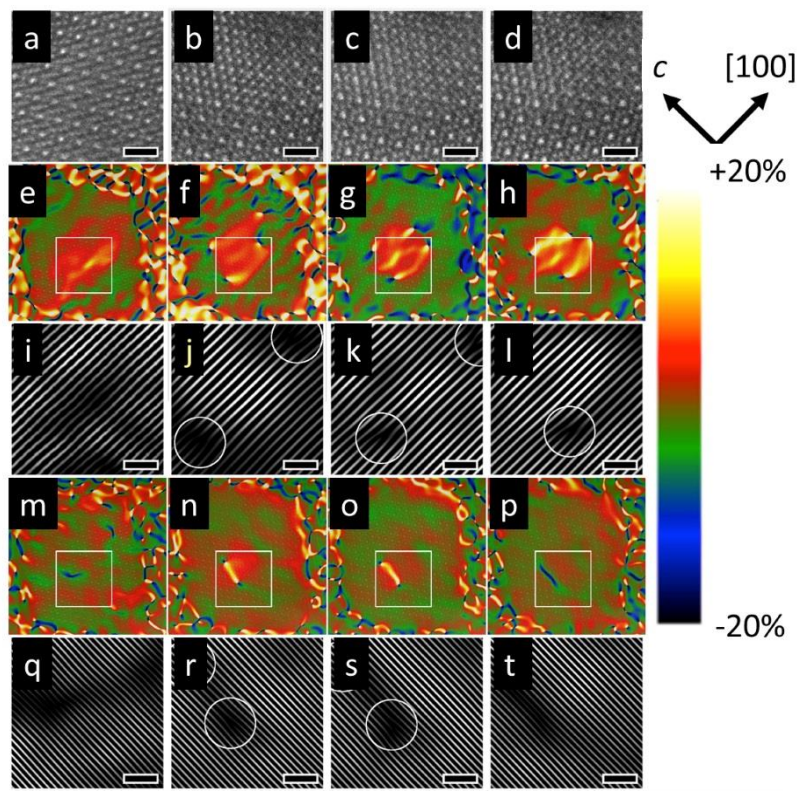

**Fig. 4.** Lattice deformation upon electron exposure in the $Mn_3O_4/Fe_3O_4$ core/shell system. (a-d) High-resolution STEM-HAADF images of the nanoparticle interface (area marked with a square in the deformation maps) at (a) initial exposure time, (b) after 30, (c) 45, and (d) 48 minutes respectively. (e-h) Deformation maps obtained by GPA using 002 MnO/ 004 $Fe_3O_4$ reflections. (i-l) Fourier-filtered images of (a-d) using 002 MnO/ 004 $Fe_3O_4$ reflections, respectively. (m-p) Deformation maps obtained by GPA using 220 MnO/ 440 $Fe_3O_4$ reflections. (q-t) Fourier-filtered images of images in (a-d) using 220 MnO/ 440 $Fe_3O_4$ reflections. The false color scale depicts the strain in %. White circles highlight misfit dislocation in the Fourier-filtered images. Scale bars represent 1 nm.

**Probing the Meta-Stability of Oxide Core/Shell Nanoparticle Systems at Atomic Resolution**

**Table of Contents**

In-situ electron-induced selective reduction of magnetic oxides is studied in core/shell nanoparticles using high-resolution electron microscopy and spectroscopy. This study highlights the relevance of understanding the local dynamics responsible for nanoscale changes to control stability and, ultimately, macroscopic functionality.

**ToC figure:**

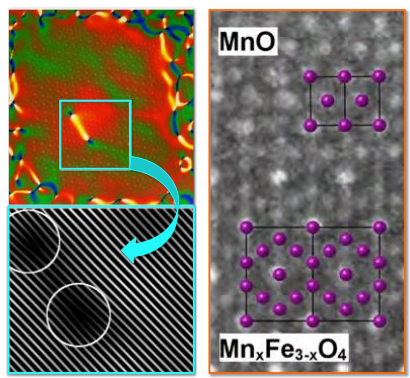

# Supplementary Information

**Probing the Meta-Stability of Oxide Core/Shell Nanoparticle Systems at Atomic Resolution**

*Manuel A. Roldan, Arnaud Mayence, Alberto López-Ortega, Ryo Ishikawa, Juan Salafranca, Marta Estrader, German Salazar-Alvarez, M. Dolors Baró, Josep Nogués,[*] Stephen J. Pennycook, Maria Varela[*]*

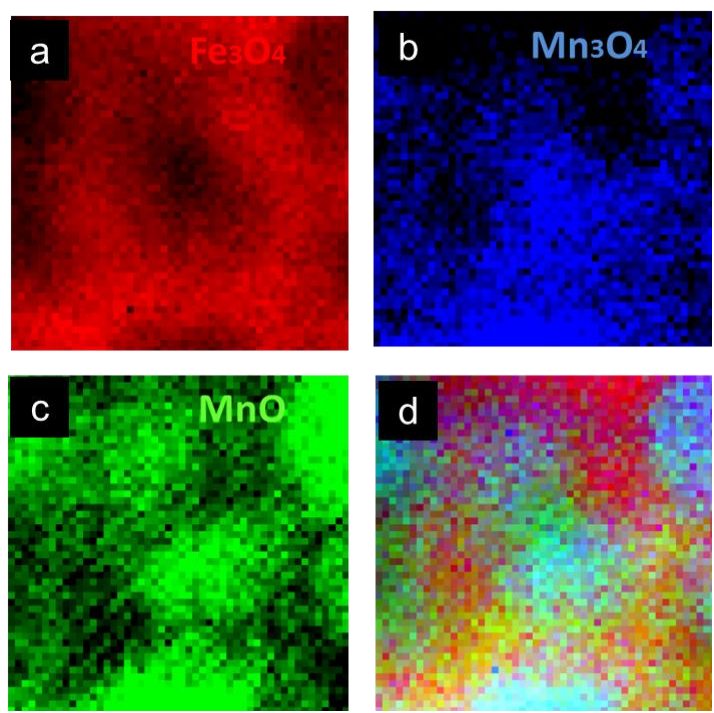

**Figure S1:.** Multiple linear least squares (MLLS) coefficient maps of a $Mn_3O_4$ /$Fe_3O_4$ nanoparticle, obtained at 200kV, for the different reference spectra: (a) $Fe_3O_4$, (b) $Mn_3O_4$ and (c) MnO, respectively. (d) Corresponding RGB colored map where the previous maps are overlaid with $Fe_3O_4$ in red, MnO in green, and $Mn_3O_4$ in blue. Image lateral size is 12nm x 12 nm.

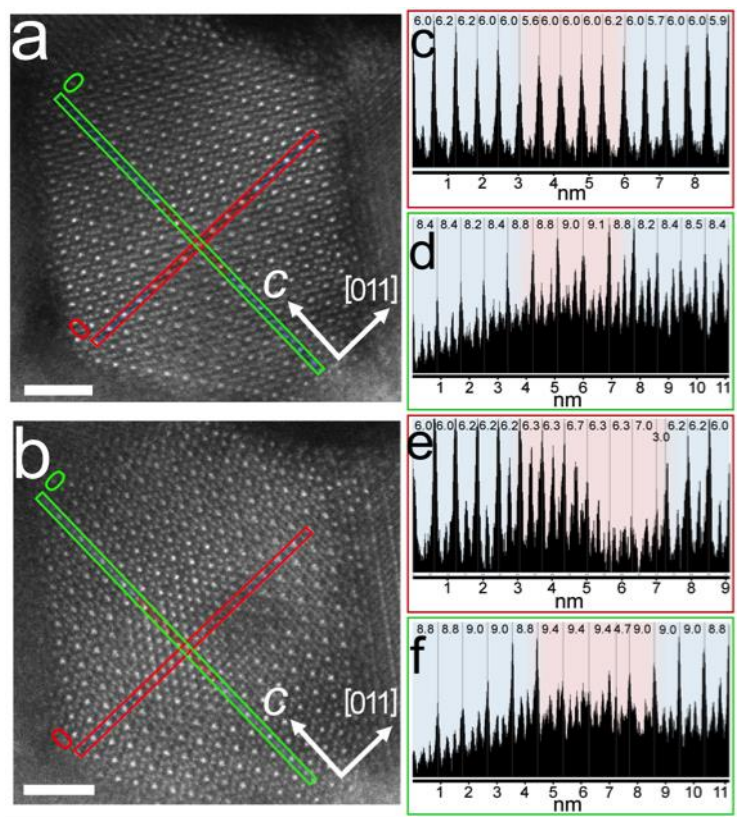

**Figure S2. Lattice spacing in Mn$_3$O$_4$|Fe$_3$O$_4$ core|shell nanoparticle at different electron exposure times**. (a, b) STEM-HAADF high resolution images of a Mn$_3$O$_4$|Fe$_3$O$_4$ core|shell nanoparticle recorded along <110> at initial exposure stage and after 30 minutes, respectively. (c, d) Line profiles of the marked regions in (a) along (c) [110] and (d) [001]. (e, f) Line profiles of the marked regions in (b) along (e) [110] and (f) [001]. Scale bar: 2nm.

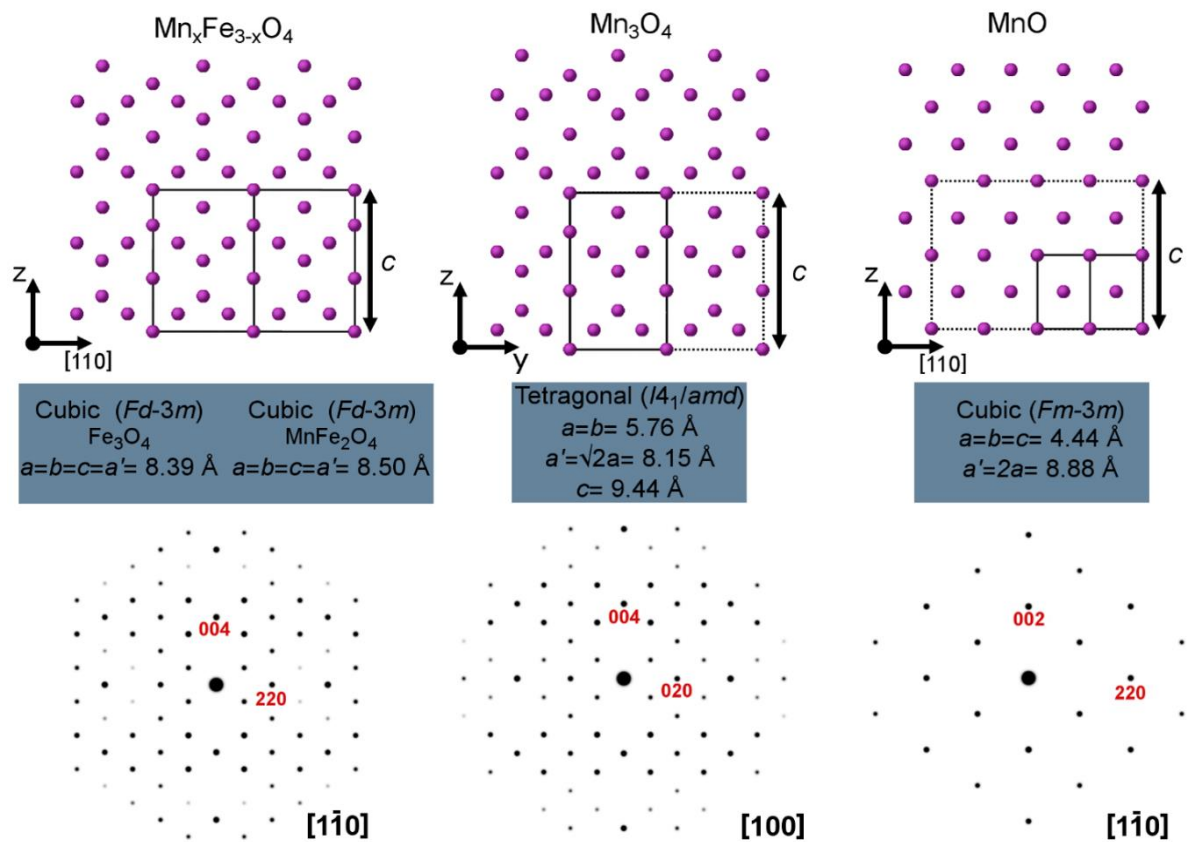

**Figure S3: Theoretical values and model for Mn$_3$O$_4$, MnO and Mn$_x$Fe$_{3-x}$O$_4$.** Spinel and rock-salt structures (top panel), with their respective simulated electron diffraction patterns (bottom panel).

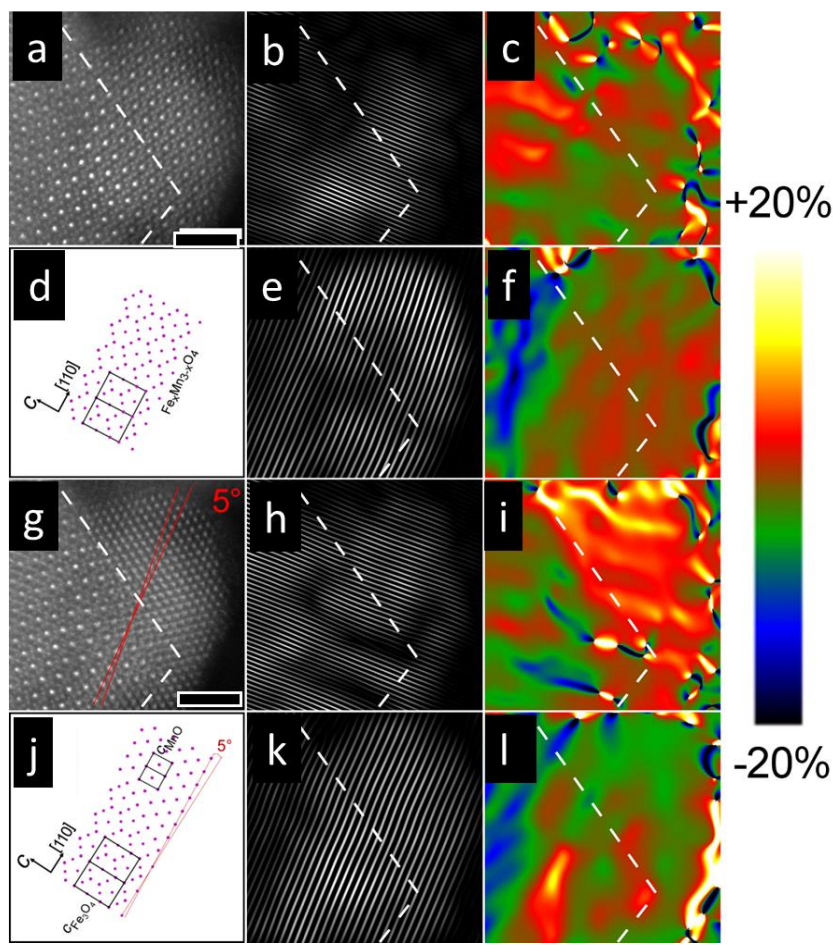

**Figure S4. GPA analysis for a Mn$_3$O$_4$/Fe$_3$O$_4$ nanoparticle.** (a) High-resolution STEM-HAADF image of the nanoparticle interface at initial exposure time. (b) Fourier filtered image of (a) using the 044$_{Fe3O4}$|022$_{MnO}$ reflection. (c) Deformation map obtained by GPA using the 044$_{Fe3O4}$|022$_{MnO}$ reflection. (d) Model of the coherent domain corresponding to a spinel structure of Mn$_x$Fe$_{3-x}$O$_4$. (e) Fourier filtered images of (a) using the 004$_{Fe3O4}$|002$_{MnO}$ reflection. (f) Deformation map obtained by GPA using the 004$_{Fe3O4}$|002$_{MnO}$ reflection. (g) High-resolution STEM-HAADF image of the nanoparticle interface after 15 minutes exposure time. (h) Fourier filtered image of (g) using the 044$_{Fe3O4}$|022$_{MnO}$ reflections. (i) Deformation map obtained by GPA using the 044$_{Fe3O4}$|022$_{MnO}$ reflection. (j) Model of Fe$_3$O$_4$|MnO domains with a 5° lattice rotation. (k) Fourier filtered images of (g) using the 004$_{Fe3O4}$|002$_{MnO}$ reflection. (l) Deformation map obtained by GPA from (k) using the 004$_{Fe3O4}$|002$_{MnO}$. The false color scale depicts the strain in %. The dashed lines inset depict the interface between Fe$_3$O$_4$ and Mn$_3$O$_4$. Scale bars: 2 nm.

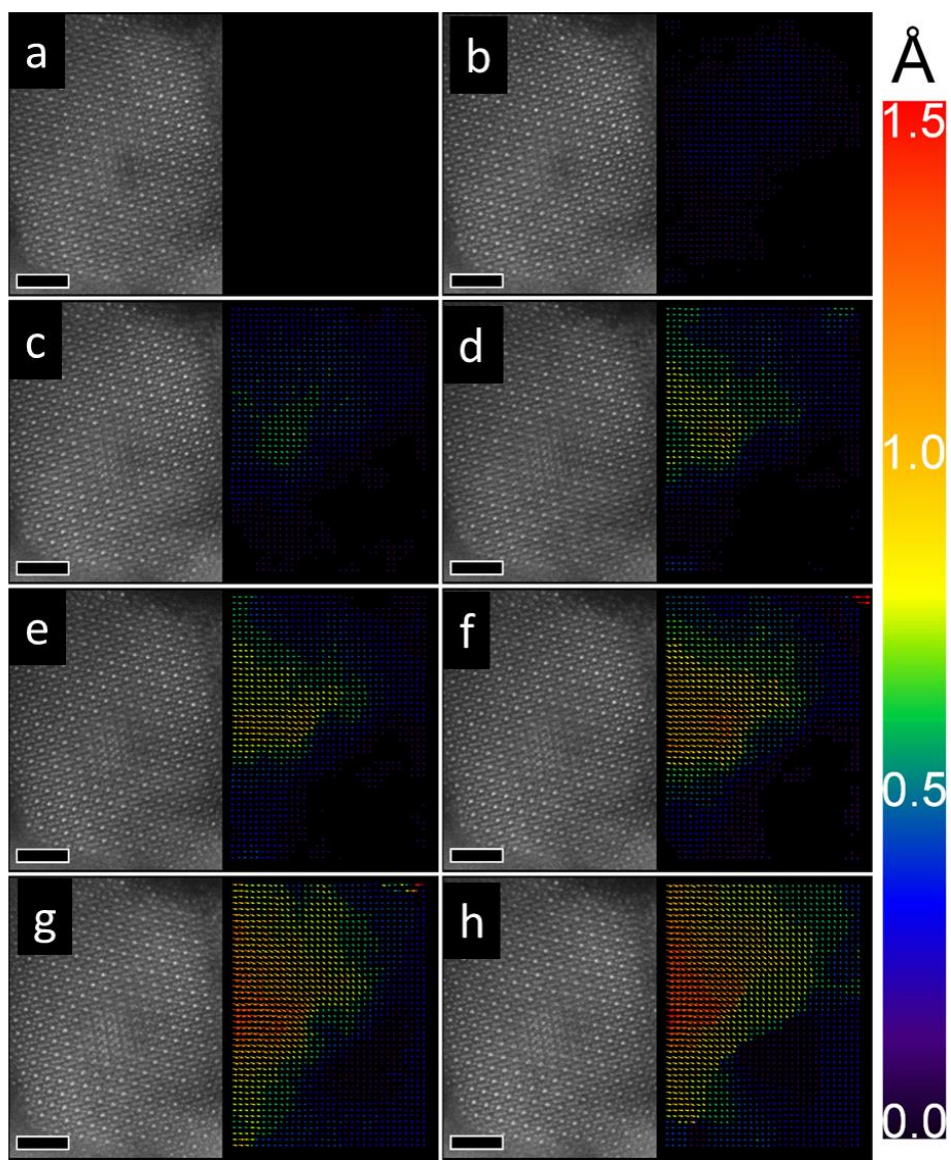

**Figure S5.** Series of subsequent high-resolution STEM-HAADF images extracted from the PIV study. The arrows in the top left panel show the crystallographic directions of the $Mn_3O_4$ structure. The scale bar represents 2 nm.

Also, **Video 1 (V1)** including the PIV images is attached as a separate file as additional supporting information.

**Thermodynamics of the $Fe_3O_4$ and $Mn_3O_4$ phases**

The electron-induced reduction of Fe(III) to Fe(II) is not observed in neither of the core|shell systems. This finding can be understood both in terms of the difference in Gibbs free enthalpy between the $Mn_3O_4 \rightarrow MnO$ and $Fe_3O_4 \rightarrow FeO$ transformation and the instability of $Fe_{1-x}O$.

**Table S1.** Bibliography Enthalpy data[1]:

| Oxide | $\Delta H_f^o$ (KJ/mol) |
|---|---|
| $Fe_3O_4$ | -1118.4 |
| FeO | -272.0 |
| MnO | -385.2 |
| $Mn_3O_4$ | -1387.8 |

Reactions involved in $Fe_3O_4|Mn_3O_4$ and $Mn_3O_4|Fe_3O_4$ core|shell nanoparticles:

- $Fe_3O_4 \rightarrow 3FeO + ½ O_2$

    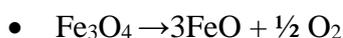  $\Delta H_r = 3 (-272.0)-(-1118.4) = 302.4$ KJ/mol $Fe_3O_4 \rightarrow 3.1$ eV       (1)

- $Mn_3O_4 \rightarrow 3MnO + ½ O_2$

    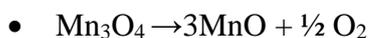  $\Delta H_r = 3 (-385.2)-(-1387.8) = 232.2$ KJ/mol $Mn_3O_4 \rightarrow 2.4$ eV       (2)

[1] Data from Chase, M. W., NIST-JANAF Thermochemical Tables, Fourth Edition J. Phys. Chem. Ref. Data Monograph 9, 1998.